\documentclass[fleqn,usenatbib]{mnras}

\usepackage{newtxtext,newtxmath}

\usepackage[T1]{fontenc}

\usepackage{lineno}

\DeclareRobustCommand{\VAN}[3]{#2}
\let\VANthebibliography\thebibliography
\def\thebibliography{\DeclareRobustCommand{\VAN}[3]{##3}\VANthebibliography}

\usepackage{graphicx}	
\usepackage{amsmath}	

\title[Multi-Zone Modeling of Blazar Jets]{Multi-zone Modeling of Blazar Jets: Constraints from GeV-Optical Correlation and Short-Timescale Variability}

\author[A. Bala, K. Mitra and R. Chatterjee]{
Arit Bala,$^{1}$
Kaustav Mitra,$^{2}$
Ritaban Chatterjee,$^{1}$\thanks{E-mail: ritaban.astro@presiuniv.ac.in}
\\
$^{1}$School of Astrophysics, Presidency University, 86/1 College Street, Kolkata - 700073, West Bengal, India\\
$^{2}$Argonne National Laboratory, 9700 S Cass Ave, Lemont, IL 60439, United States
}

\date{Accepted XXX. Received YYY; in original form ZZZ}

\pubyear{\the\year{}}

\begin{document}
\label{firstpage}
\pagerange{\pageref{firstpage}--\pageref{lastpage}}
\maketitle

\begin{abstract}
We have developed a multi-zone model of blazar jet emission, in which the emission region contains many cells with individual magnetic fields and electron energy distributions. Nonthermal emission from radio to $\gamma$-rays is generated by electrons accelerated by shocks passing through the region via synchrotron and inverse-Compton (IC) processes. The optical and GeV variability at days-to-months time-scale simulated from our model are strongly correlated with no significant time lag, as observed in most blazars and indicated by the standard shock-in-jet model. However, the mechanism of the shorter time-scale variability has been less explored, although such fluctuations at X-rays, $\gamma$-rays and optical bands have been observed regularly in recent years. In our model, the hr time-scale variability of the synchrotron radiation is due to the spatial fluctuation of the magnetic field in the emission region. We found that to reproduce the short-timescale variability of the observed synchrotron emission in blazars, the required fluctuations of the magnetic field are in the range $1-2\%$ to $25-30\%$. Similar variability of the IC emission, which does not depend on the magnetic field, may be reproduced in our model by implementing equipartition of energy between the magnetic field and particles. We found that orphan flares in the optical or GeV band, or optical-GeV correlation with a significant time delay, as observed occasionally, may be reproduced in certain special conditions related to the orientation of the magnetic field in the cells. 
\end{abstract}




\section{Introduction} \label{sec:intro}

Blazars are a class of radio-loud active galactic nuclei (AGN), in which the relativistic jets are pointed close to the line of sight of the observer \citep{Blandford.Rees.1978b, Blandford.et.al.2019, Talvikki.Hovatta.2019}. Jets of blazars extend up to hundreds of kpc \citep{Celotti.et.al.2001, Tavecchio.et.al.2007, Marscher.Jorstad.2011} from the central engine, and they emit throughout the electromagnetic spectrum, often from radio to $\gamma$-rays \citep{Madejski.Sikora.2016, Markus.Bottcher.2019} reaching, in some cases, TeV energies \citep{Punch.et.al.1992, Wakely.Horan.2008}. 

A key feature of blazars is the double-hump structured spectral energy distribution \citep[SED;][]{Fossati.et.al.1998}. Modeling suggests that the lower energy hump of the SED is generated due to synchrotron radiation from the relativistic electrons present in the jet \citep{Joel.N.Bregman.1981, C.M.Urry.1982, C.D.Impey.1988, G.Ghisellini.2017}. The higher-energy hump is produced by inverse-Compton (IC) radiation from the same electron population. The IC process can be subdivided into two classes depending upon the source of the seed photons. When the photons produced due to the synchrotron process in the jet itself get up-scattered by the electrons in the jet, it is called the synchrotron self-Compton (SSC) process \citep{A.Konigl.1981, L.Maraschi.1992, M.Sikora.1994, Bloom.&.Marscher.1996, A.Mastichiadis.1997, Markus.Bottcher.2007}. On the other hand, if the seed photons come from outside the jet, e.g., the dusty torus \citep{Marek.Sikora.2008} or the broad line region (BLR) \citep{G.Ghisellini.1998}, it is known as the external Compton (EC) mechanism \citep{C.D.Dermer.1992, M.Blazejowski.2000, Markus.Bottcher.2010}. The above scenario is commonly referred to as the leptonic mechanism of emission. The higher-energy hump of blazar SEDs may also be modeled using proton synchrotron and proton-photon cascade, commonly known as the hadronic scenario \citep{Mucke.A.2001, Mucke.A.2003, Markus.Bottcher.2013, M.Ackerman.2016}. For this work, we are considering only the leptonic model. 

Strong emission lines are observed in some blazars, particularly during intervals when the jet emission is weak \citep{Stocke.et.al.1991, Landt.et.al.2004, Fan.J.H.2006, Padovani.2007}. Those are called the flat spectrum radio quasars (FSRQs). In contrast, the so-called BL Lac-type blazars do not exhibit prominent emission lines. It is believed that in the latter, the luminosity of the accretion disk is itself low \citep{Celotti.et.al.1997}. As a result, the BLR \citep{G.Ghisellini.1998} does not generate strong emission lines and emission from the dusty torus \citep{M.Blazejowski.2000, Marek.Sikora.2008} is also relatively weak. Thus, the main inverse-Compton process for the BL Lac-type blazars is considered to be SSC. 

Apart from the above classification, blazars have been divided into three subgroups based on the position of the synchrotron peak in their SED. If the synchrotron peak is at the IR or optical regime, it is known as low synchrotron-peaked (LSP); if the peak is at the X-ray frequencies, the blazar is called high synchrotron-peaked (HSP), and any blazar with the synchrotron peak in between is called intermediate synchrotron-peaked (ISP) blazar \citep{Padovani.Giommi.1995, Abdo.2010.a}. FSRQs are usually LSP-type (in a few cases ISP type) blazars with a luminous accretion disk, while BL Lacs may be LSP, ISP or HSP-type blazars \citep{Urry.Padovani.1995, G.Ghisellini.1998, Abdo.2010.a, Abdo.2010.b, G.Ghisellini.2017}. 

As the relativistic jets of blazars are pointed towards the observer, due to Doppler boosting effect the flux measured in the observer's frame increases by a factor of $\sim$$\delta^4$ and the variability time-scale decreases by a factor $\delta$, where $\delta$ is the Doppler factor and its value is usually in the range of a few to a few tens \citep[e.g.,][]{jorstad2005polarimetric, lio18, hov09, rajguru22}. Hence, fast and high-amplitude variability at multiple wave bands is one of the most noticeable attributes of blazars \citep{Aharonian_F.et.al.2007, Albert_J.2007, Ackermann.et.al.2011, Arlen_T.2013, P.Penil.2024}. It has been observed to be present at longer (months, years, decades) to shorter (weeks to days) time-scales \citep{Bonning.et.al.2012, Ritaban.Chatterjee.2012, Majumder.et.al.2019} at radio to GeV bands. At X-ray frequencies, blazar variability has also been monitored at timescales shorter than above, e.g., hours to minutes by different telescopes including, most recently, XMM-Newton and AstroSat \citep[e.g.,][]{Bhattacharyya.et.al.2020, Susmita.Das.2023} while Kepler and Transiting Exoplanet Survey Satellite (TESS) have observed the same at optical wavelengths  \citep{Raiteri.et.al.2021, pininti.et.al.2023}. 

Fermi Gamma-Ray Space Telescope, which was launched in 2008, has detected more than 3000 blazars \citep{Abdollahi.et.al.2020, Abdollahi.et.al.2022}. A few hundred of those have been monitored at radio, optical and X-ray wave bands by other observatories simultaneously with Fermi, particularly when those were bright at GeV energies \citep[e.g.,][]{Ackermann.et.al.2011, P.Penil.2024}. Many authors have carried out cross-correlation studies of the GeV variability of blazars obtained from Fermi-LAT with that at longer wavelengths. Various properties of blazar emission have been determined from the strength and time delay of the above correlations. For example, the above explanation of the two-humped nature of the blazar SED, e.g., optical and GeV emission generated by the same electron distribution in the jet via synchrotron and IC processes, has been confirmed from the strong GeV/optical correlation seen in many blazars \citep{Hovatta.et.al.2014, Cohen.et.al.2014, Liodakis.et.al.2019, Rajput.et.al.2020}; monitoring with ground- and space-based telescopes as well as with very long baseline radio interferometry has been used to constrain the location of the nonthermal emission region, in some cases, to be a few to tens of pc down the jet from the central engine \citep{nalewajko.et.al.2014, Costamante.et.al.2018, Barat.et.al.2022, Kundu.et.al.2025} and to establish the relation between multi-wavelength outbursts and radio-bright knots moving down the jet as observed in VLBA images \citep{Marscher.et.al.2008, Marscher.et.al.2010, Agudo.et.al.2011}. 

However, several aspects of blazar emission are not well understood. For example, while the variability of emission at the two peaks of the SED, e.g., optical and GeV are often strongly correlated, sometimes the correlation is weak or absent \citep{Liodakis.et.al.2019, Rajput.et.al.2020}. While the long-term (weeks to months) variation is supposedly caused by the motion of shocks down the jet and resulting energization of the emitting particles, the mechanism of the shorter timescale (hours to days) variability is uncertain. The power spectral density of blazar emission variability at multiple wave bands is often consistent with a simple power-law \citep{Abdo.2010.b, Ritaban.Chatterjee.2012, Bhatta.Dhital.2020, Arti.Goyal.2020, Arti.Goyal.2021, Shah.et.al.2025} although the inferred scale-invariant nature of the amplitude of variability at a long range of timescales (hours to years) has not been explained in the literature. Therefore, a systematic approach is required to draw robust inferences from a large database of blazar variability information as has been accumulated by Fermi and other multi-wavelength initiatives over the last 15 years or more. Modeling the SED of blazars is another prominent method to constrain the relevant parameters of blazar emission, but there is usually a large amount of degeneracy among the best-fit parameters.

In this work, we develop a multi-zone numerical model to simulate blazar emission variability at multiple wave bands. Multi-zone models of blazar emission have been developed and used by many authors \citep[e.g.,][]{Graff.et.al.2008, Joshi.Bottcher.2011, Alan.P.Marscher.2014, Chen.et.al.2014, Chen.Bottcher.2015, Liu.et.al.2023, Xin.yu.hu.2024}. Those models focus on different aspects of jets such as its geometry or relevant acceleration and emission mechanisms based on the specific goal of the work. In our model, we focus on multi-wavelength variability at a range of time-scales in order to compare those with the large sample of observed multi-band blazar light curves, including those at sub-day time-scales. See \citet{A.Kundu.et.al.22, Das.Chatterjee.25, Kundu.et.al.2025} for the description and application of previous, more simplified, versions of our model. We compare the properties of the simulated light curves and their cross-correlation results to those of the observed data in order to constrain the possible parameter space. Our model has been described in detail in \S\ref{sec:model}. \S\ref{sec:results} describes the results obtained from our model, followed by comparison with observations in \S\ref{sec:compare}. Finally, \S\ref{sec:discussion} contains a summary of the results obtained and related discussion.

\section{Theoretical Modeling} \label{sec:model}

\begin{figure*}
\includegraphics[width=\textwidth]{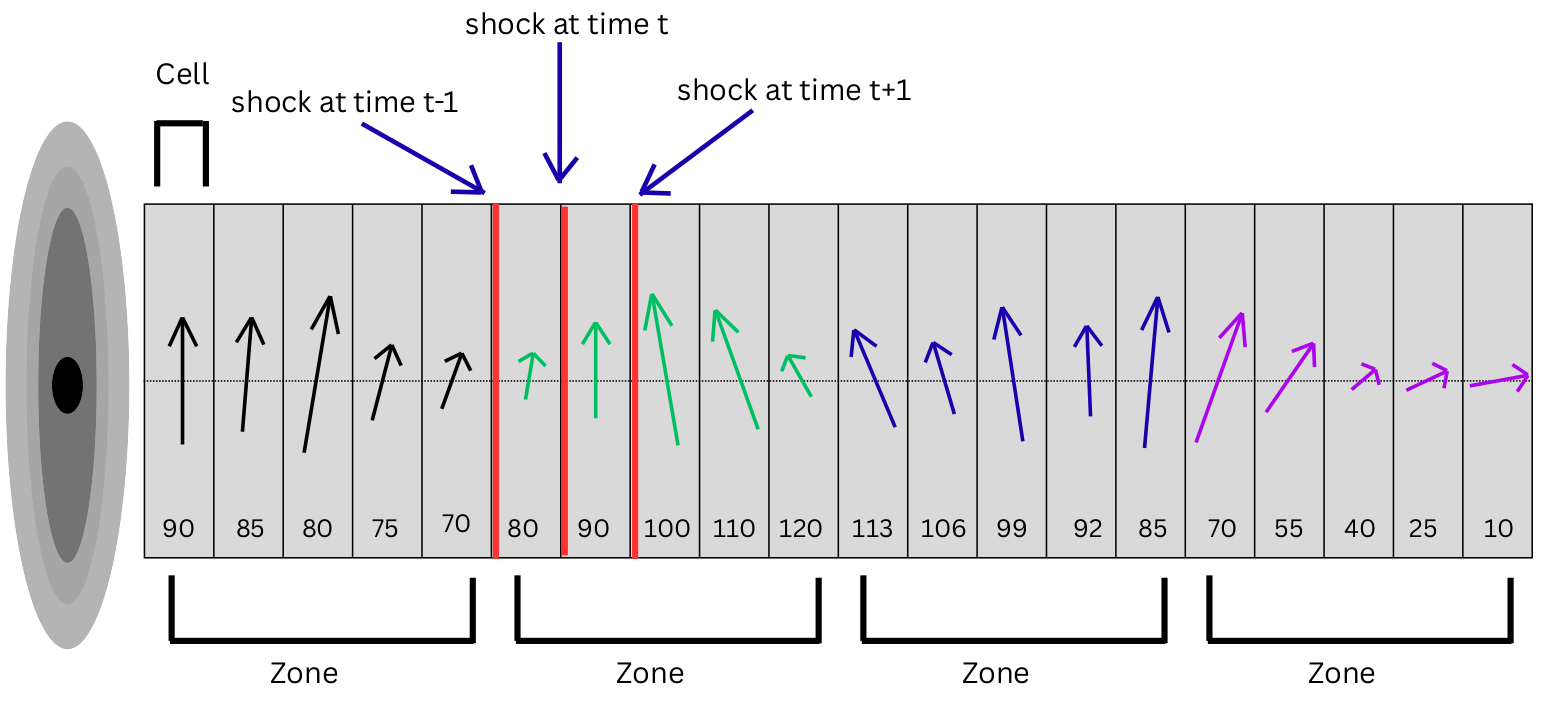}
\caption{A schematic description of the geometry of our model. Each cell of the emission region has a different magnetic field value and direction. The values shown at the bottom of the cells indicate the angle of the magnetic field with the jet axis. In a zone, the difference in the angle between one cell and its adjacent cells is constant. Different zones have been indicated using a different color scheme of arrows. The lengths of the arrows represent the values of the magnetic field.
\label{fig:Model_fig}}
\end{figure*}

\begin{table*} 
\centering
\caption{Range of input parameter values used in this study}
\label{tab:parameter_space}
\begin{tabular}{lccr}
\hline
\hline
parameters & minimum value & maximum value & most used value \\
\hline
\hline
B field (BL Lac) & 0.01 Gauss & 0.8 Gauss & 0.08-0.2 Gauss \\
$\gamma_{min}$ (BL Lac) & 2 & 100 & 10 \\
$\gamma_{max}$ (BL Lac) & $3\times 10^3$ & $6\times 10^5$ & depends on the location of synchrotron peak \\
B field (FSRQ) & 0.5 Gauss & 10 Gauss & 0.8-1.2 Gauss \\
$\gamma_{min}$ (FSRQ) & 2 & 100 & 10 \\
$\gamma_{max}$ (FSRQ) & $1\times 10^3$ & $5\times 10^3$ & $3\times 10^3$ \\
accretion disk luminosity & $1.2\times 10^{44}$ erg\,s$^{-1}$ & $1.2\times 10^{46}$ erg\,s$^{-1}$ & $1.2\times 10^{46}$ erg\,s$^{-1}$ \\
\hline
\end{tabular}
\end{table*}

\begin{table*} 
\centering
\caption{Fixed input parameter values used in this study}
\label{tab:parameter_space_fixed}
\begin{tabular}{lcr}
\hline
\hline
fixed parameters & fixed value & explanation \\
\hline
\hline
power law index (s) & 2.05 & for mildly relativistic shock $s\gtrsim2$ \citep{Longair.book.2011} \\
Ratio of magnetic to relativistic electron energy density ($f_b$) & 1 & we use equipatition of energy \\
Bulk Lorentz factor ($\Gamma$) & 15 & \\
Angle between jet axis and line of sight & $0.05^r$ & \\
Doppler factor ($\delta$) & 20 & calculated using the above two parameters \\
Fraction of reprocessing of disk photons for BLR & 0.1 & value from literature \citep{Hayashida.et.al.2012}\\
Fraction of reprocessing of disk photons for torus & 0.01 & value from literature \citep{Hayashida.et.al.2012}\\
Radius of jet & 0.001 pc &  \\
\hline
\end{tabular}
\end{table*}

\subsection{Emission Region} \label{subsec:region}

We consider a cylindrically shaped emission region in the jet, which we divide into multiple `cells' by slicing along the axis (a pictorial representation can be found in Figure \ref{fig:Model_fig}). A mildly relativistic shock ($\beta_{shock}=0.2$ w.r.t. the emission region; $\Gamma_{shock}\approx1.02$) passes through the emission region and energizes the initially quiescent electrons to a simple power-law energy distribution, after which the electrons start emitting through synchrotron and inverse-Compton processes. This is a `multi-zone' model in the sense that the cells in the emission region may have different magnetic fields and electron energy distributions, the latter of which evolves with time independent of the other cells due to radiative cooling. We generate light curves and SED due to the above non-thermal processes in order to compare with the observed data.

In the convention of our model, the shock front takes exactly one time step to move from one cell to the next. Therefore, the cell size is $v_{shock}*dt$, where $v_{shock}$ is the speed of the shock front and the smallest time step is $dt$. The total length of the emission region depends on the number of cells. 

\subsection{Magnetic Field} \label{subsec:B}

In our model, there is spatial variation of the magnetic field in the emission region but in a given cell the magnetic field is constant over time. The magnitude of the magnetic field is given by $B = B_1 + B_2$, where $B_1$ and $B_2$ are a smoothly varying and a fluctuating component, respectively. The values of $B_1$ at the first and last cell ($B_{1,ini}$ and $B_{1,final}$, respectively) are given as input parameters. $B_1$ varies linearly from $B_{1i}$ to $B_{1f}$. We split the emission region into $N$ cells divided into $k$ `zones'. The zones are defined such that the magnetic field within each zone is somewhat correlated. The fluctuating component, $B_2$, is drawn from a uniform random distribution normalized by a constant factor in order to control the relative contribution of the fluctuating magnetic field w.r.t. the smoothly varying component. Similarly, additional fluctuations are introduced in the cells within a zone, e.g., a similar random distribution but normalized by a different constant such that the fluctuations within a zone and between zones may be independently characterized. In addition, the direction of the magnetic field may also change from cell to cell, which is controlled by another parameter $\delta\theta$. We generate a random angle and assign that to the first cell in a zone, and the angle varies within the zone as,

\begin{equation}
\theta_{i+1} = \theta_i + \delta\theta
\end{equation}
In order to ensure that the direction does not suddenly change by a large amount, we assign the direction of the last cell of the previous zone to the first cell of the next zone, i.e.,

\begin{equation}
\theta_i(new-zone) = \theta_f(previous-zone) + \delta \theta',
\end{equation}
where $\delta \theta'$ is another angle drawn from a random distribution. The components of the magnetic field perpendicular and parallel to the observer's line of sight in the $i$-th cell are given by:
\begin{equation}
B_{\parallel}(i) = (B_1(i)+B_2(i)) \cos\theta(i)
\end{equation}
and
\begin{equation}
B_{\perp}(i) = (B_1(i)+B_2(i)) \sin\theta(i)
\end{equation}

\subsection{Acceleration and Energy Distribution of Electrons} \label{subsec:distribution}

We assume that a shock front passes through the emission region, and when it reaches a given cell, the electron energy distribution in it, expressed in terms of the electron Lorentz factor ($\gamma$), is given by the following within the range $\gamma_{min}$ to $\gamma_{max}$:
\begin{equation}
N(\gamma) = N_0 \gamma^{-s},
\end{equation}
where $N_0$ and $s$ are constants. For this work we used $s =$ 2.05, which is consistent with the acceleration of particles by a mildly relativistic shock. We use the equipartition theorem to relate the electron energy density and the magnetic field as given below:
\begin{equation}
\frac{B^2}{8\pi} = \int_{\gamma_{min}}^{\gamma_{max}} \gamma m_e c^2 N_0 \gamma^{-s} d\gamma
\end{equation}

\begin{equation}
N_0=\frac{B^2}{8\pi m_e c^2} \left[\frac{s-2}{\gamma_{min}^{2-s}-\gamma_{max}^{2-s}}\right].
\end{equation}
The equipartition condition between magnetic field ($U_{\rm B}$) and electron energy density ($U_{\rm e}$) is invoked at the moment of particle injection when a shock reaches any particular cell. Several authors have advocated for this type of equipartition: (a) on the basis of minimum energy or minimum power configuration \citep{Duran.et.al.2013, Zdziarski.2014, Dermer.et.al.2014}, (b) using detailed particle-in-cell simulation to show that magnetic reconnection leads to $U_{\rm B}/U_{\rm e} \sim 1$ equipartition \citep{Sironi.et.al.2015, Petropoulou.et.al.2016, Christie.et.al.2019}, and (c) by invoking the assumption in theoretical models, the results of which are consistent with observational constraints \citep{Potter.et.al.2012, Potter.et.al.2013, Yan.et.al.2015, Anjum.et.al.2025, Lainez.et.al.2025}.

\subsection{Cooling of Electrons} \label{subsec:cooling}

In the presence of a magnetic field, the relativistic electrons in blazar jets lose energy via synchrotron radiation. In addition to the synchrotron photons thus produced, external photons, e.g., from the BLR and dusty torus, are present in the jet. Therefore, the high-energy electrons also cool through the inverse-Compton (IC) scattering of those photons to higher energies \citep[e.g.,][]{ryb79,Alan.P.Marscher.2014}. The energy of electron and number density distribution change due to radiative cooling as follows \citep{Longair.book.2011}:
\begin{equation}
    \gamma_f = \left[\gamma_i^{-1} + k_r\left(B^2+8\pi\left(U_{BLR}+U_{tor}+U_{SSC-in}\right)\right)\right]^{-1}
\end{equation}
\begin{equation}
    N(\gamma,t) = N_0\gamma^{-s}\left(1-\gamma kt\right)^{s-2},
\end{equation}
where $k_r = \frac{1}{8\pi}\left(\frac{4}{3}\frac{\sigma_T c}{m_e c^2}\right)$, $\sigma_T$ being the Thompson scattering cross-section and $m_e$ is electron mass, $N_0$ is calculated using Equation 7 and $k=k_r[B^2+8\pi\left(U_{BLR}+U_{tor}+U_{SSC-in}\right)]$. The electron distribution is assumed to be isotropic in the comoving frame of the emission region.

\subsubsection{Synchrotron Radiation} \label{subsubsec:synchrotron}

We calculate the synchrotron emission coefficient for each cell at each time step by the equation,

\begin{equation}
j^s_\nu (\nu) = \frac{\sqrt{3} e^3}{4 \pi m_e c^2} B sin \theta \int_{\gamma_{min}}^{\gamma_{max}} N(\gamma) F(\nu/\nu_c) d\gamma,
\end{equation}

\noindent where $\theta$ is the angle between the magnetic field $\vec{B}$ and the line of sight, $\nu _c = 3eBsin\theta \gamma^2 / 4 \pi m_e c$ and $F(\nu / \nu_c)$ is given by
\begin{equation}
F(\nu/\nu _c) = \frac{\nu}{\nu _c} \int_{\frac{\nu}{\nu _c}}^{\infty} K_{(5/3)} (\xi) d\xi 
\end{equation}
\noindent $k_{(5/3)} (\xi)$ is the modified Bessel function of the 5/3rd order. From this emission coefficient, we then calculate the total synchrotron emission at each time step.

\subsubsection{Inverse-Compton (IC) Process} \label{subsubsec:IC}

The photons generated by the synchrotron process and photons from sources external to the jet get up-scattered by the same electron population, giving rise to the higher-energy hump in the SED. The inverse-Compton emissivity (erg\,cm$^{-3}$s$^{-1}$Hz$^{-1}$sterad$^{-1}$) is given by
\begin{equation}
j_{IC}(\nu) = \int_{\gamma_{min}}^{\gamma_{max}}d\gamma N(\gamma) \int d\nu _i I_{seed}(\nu_i) \sigma _{IC} (\nu , \nu_i , \gamma),
\end{equation}
\noindent where $I_{seed}$ is the seed photon intensity and $\sigma_{IC}$ is the scattering cross-section for the inverse-Compton process. Here, $I_{seed}$ is in the jet co-moving frame, $\sigma_{IC}=\frac{3\sigma_T x}{32\nu_i}\left[8+2x-x^2+4xln\left(\frac{x}{4}\right)\right]$ \citep{Blumenthal.Gould.1970}, and $x=\nu/(\nu_i\gamma^2)$. Using those synchrotron and IC emissivities, we then calculate the SEDs and light curves.

We assume that the emission region is optically thin over the wavelength ranges considered here \citep[e.g.,][]{Joshi.Bottcher.2011}; hence, we do not include absorption/opacity effects between zones. For SSC, each cell receives synchrotron seed photons from other cells with the expected $1/r^{2}$ dependence, so contributions from distant cells are negligible. We treat the entire emission region as a single co-moving frame, so there is no relative motion between zones, and we do not apply additional beaming or aberration effects that would introduce anisotropy within the emission region. Synchrotron self-absorption (SSA) has not been included in our model because we are primarily interested in the GeV and optical variability, and SSA becomes significant at longer wavelengths \citep[e.g.,][]{Finke.Becker.2014, Liu.et.al.2023}.

\subsection{Other Parameters} \label{subsec:others}

\begin{figure*}
\includegraphics[width=\textwidth]{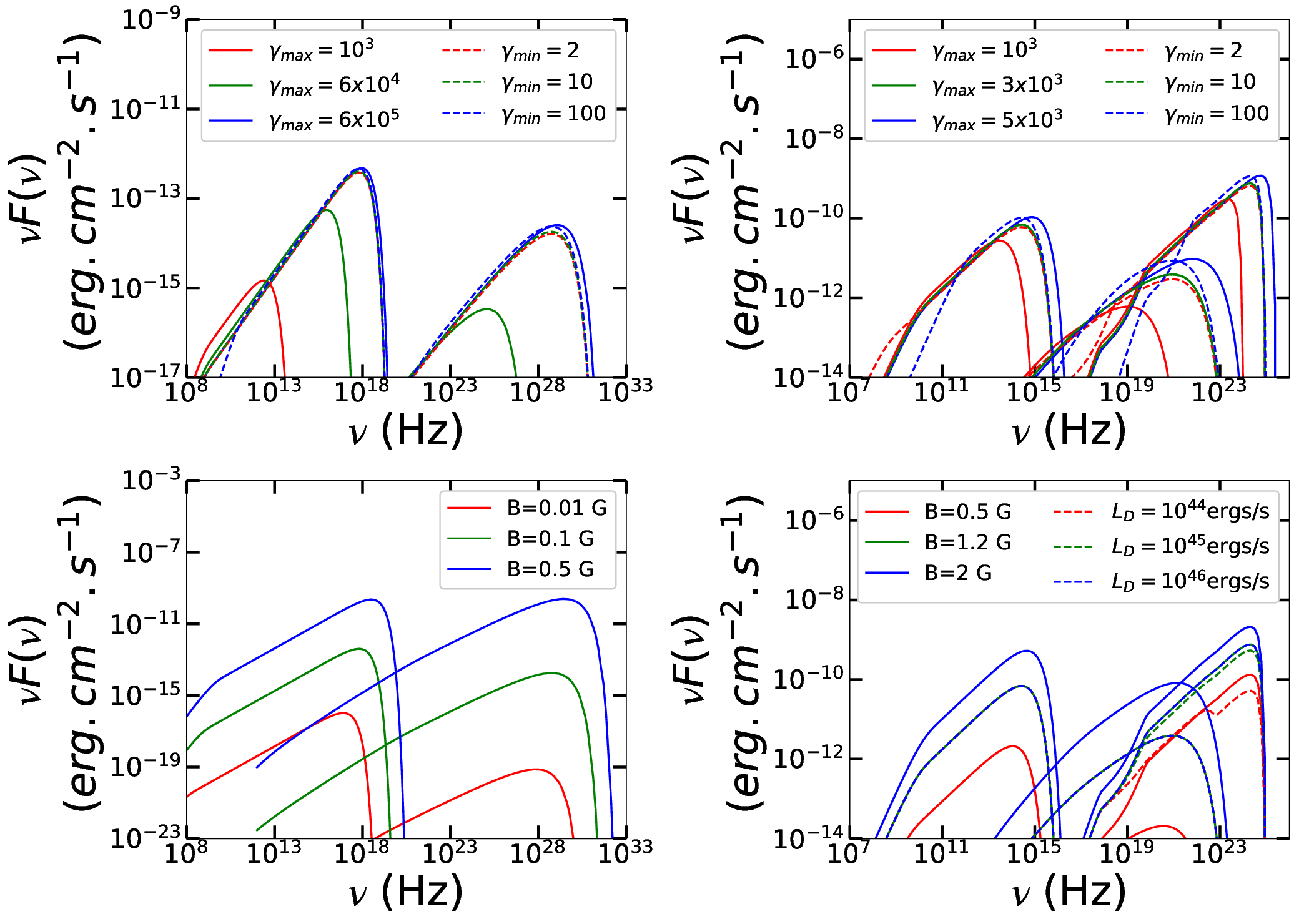}
\caption{SEDs of blazars with different sets of values of the relevant input parameters. 
\textbf{Top left} panel shows the SED of BL Lac type blazars by varying $\gamma_{max}$ and $\gamma_{min}$ (constant values used for parameters are $\gamma_{min}=2$ in the case of $\gamma_{max}$ variation, $\gamma_{max}=6\times10^5$ in case of $\gamma_{min}$ variation, $B\approx0.1$ Gauss).
\textbf{Top right} panel shows the SED of FSRQs by varying $\gamma_{max}$ and $\gamma_{min}$ (constant values used for parameters are $\gamma_{min}=2$ in the case of $\gamma_{max}$ variation, $\gamma_{max}=3\times10^3$ in case of $\gamma_{min}$ variation, $B\approx1.2$ Gauss, $L_D=10^{46}$ ergs \, s$^{-1}$).
\textbf{Bottom left} panel shows the effect of varying the magnetic field on the SEDs of BL Lac type blazars (constant values used for the parameters are $\gamma_{min}=2$, $\gamma_{max}=6\times10^5$).
\textbf{Bottom right} panel shows the effect of varying the magnetic field and accretion disk luminosity on the SEDs of FSRQs (constant values used for parameters are $\gamma_{min}=2$, $\gamma_{max}=3\times10^3$, $B\approx1.2$ Gauss in case of $L_D$ variation, $L_D=10^{46}$ ergs \, s$^{-1}$ in case of $B$ variation).
\label{fig:varying_SEDs}}
\end{figure*}

We calculate the light curves and SEDs in the jet frame and use the Doppler factor ($\delta$) to compute the same in the observer's frame in order to compare with observed data. The Doppler factor is determined from the bulk Lorentz factor ($\Gamma$) and the viewing angle ($\Theta$) between the observer's line of sight and the jet axis, which are given as input parameters of the model.

Other important parameters for our model are the distance of the emission region from the central engine and the luminosity of the disk. We consider external radiation from the BLR and torus. Accretion disk emission is assumed to be de-beamed in the jet frame and hence negligible. If the distance of the emission region from the central engine is comparable to that of the BLR/torus, we consider that external radiation to be isotropic. If not, relativistic beaming is taken into account. We assume the BLR and torus emission to be approximately monochromatic with frequencies $10^{16}$ and $10^{14}$ Hz, respectively, with a log-normal spectrum having a small width around the above frequencies such that the effect of the shape of the spectra on the results is negligible. The BLR and torus luminosities are necessary for computing the seed photon density in the emission region for IC scattering. The BLR and torus energy densities are calculated using the following equations \citep{Hayashida.et.al.2012}:
\begin{equation}
    U_{BLR}(r)=\frac{\epsilon_{BLR}\Gamma^2L_D}{3\pi r_{BLR}^2c\left[1+(r/r_{BLR})^{\beta_{BLR}}\right]}
\end{equation}
\begin{equation}
    U_{torus}(r)=\frac{\epsilon_{torus}\Gamma^2L_D}{3\pi r_{torus}^2c\left[1+(r/r_{torus})^{\beta_{torus}}\right]},
\end{equation}
where $r$, $\Gamma$, and $L_D$ are the distance of the cell from the central engine, the bulk Lorentz factor of the jet, and the luminosity of the accretion disk, respectively. We use the scaling relations $r_{BLR}=0.1(L_D/10^{46})^{0.5}$pc, $r_{torus}=2.5(L_D/10^{46})^{0.5}$pc for the distances of the BLR and torus from the central engine, where $L_D$ is given in ergs/s. $\epsilon_{BLR}=0.1$, $\epsilon_{torus}=0.01$ are the fractions of disk luminosity reprocessed by the BLR and torus, respectively. We take the values of the parameters $\beta_{BLR}=3$, $\beta_{torus}=4$ \citep{Hayashida.et.al.2012}.

\section{Simulation Results} \label{sec:results}

\begin{figure*}
\includegraphics[width=\textwidth]{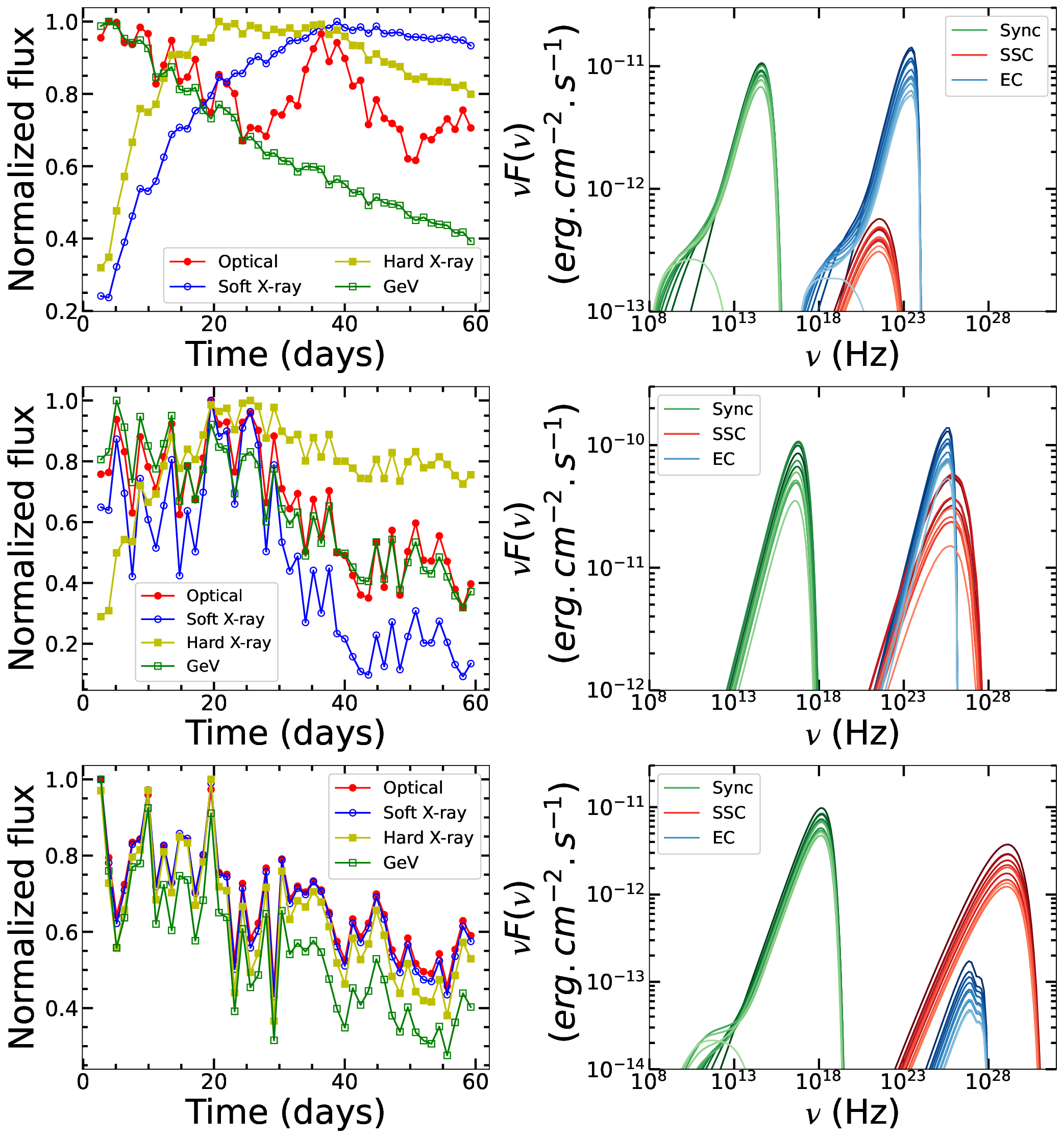}
\caption{Normalized optical, soft X-ray, hard X-ray and $\gamma$-ray light curves and SEDs, for an LSP type blazar (\textbf{top panels}), an ISP type blazar (\textbf{middle panels}) and an HSP type blazar (\textbf{bottom panels}). The right panels show the time evolution of the SEDs with darker shades indicating earlier times.
\label{fig:LC_SED_all_type}}
\end{figure*}

To study how various parameters affect the multi-wavelength emission, we generate SEDs for different combinations of the magnetic field ($B$), $\gamma_{max}$, $\gamma_{min}$ and accretion disk luminosity ($L_{disk}$). We use ranges of values of the above parameters obtained from the literature \citep[e.g.,][]{Ghisellini.et.al.2010,Castignani.et.al.2017,Dutka.et.al.2017,Paliya.et.al.2017}, which have been shown in Table \ref{tab:parameter_space}. Different sets of values are used for the FSRQ and BL Lac-type blazars. 

Figure \ref{fig:varying_SEDs} exhibits the effect of $\gamma_{min}$, $\gamma_{max}$, $B$, and $L_{disk}$ on the simulated SEDs. We find that a change in the value of $\gamma_{min}$ from 2 to 100 has little effect on the SEDs, while $\gamma_{max}$ and $B$ cause significant change in the synchrotron peak and the total emission, as expected. The other parameter, e.g. the luminosity of the accretion disk, has an impact on the EC radiation, but it does not affect the synchrotron or SSC directly. However, if EC cooling of electrons, e.g., increases, in turn, the synchrotron and SSC cooling may decrease. Therefore, the accretion disk luminosity indirectly affects the synchrotron and SSC, although that effect may be small. Figure \ref{fig:LC_SED_all_type} shows light curves and the corresponding SEDs simulated from our model with a range of input parameters. Those are discussed in the next section.

\section{Comparison With Observation} \label{sec:compare}

\begin{figure}
\includegraphics[width=\columnwidth]{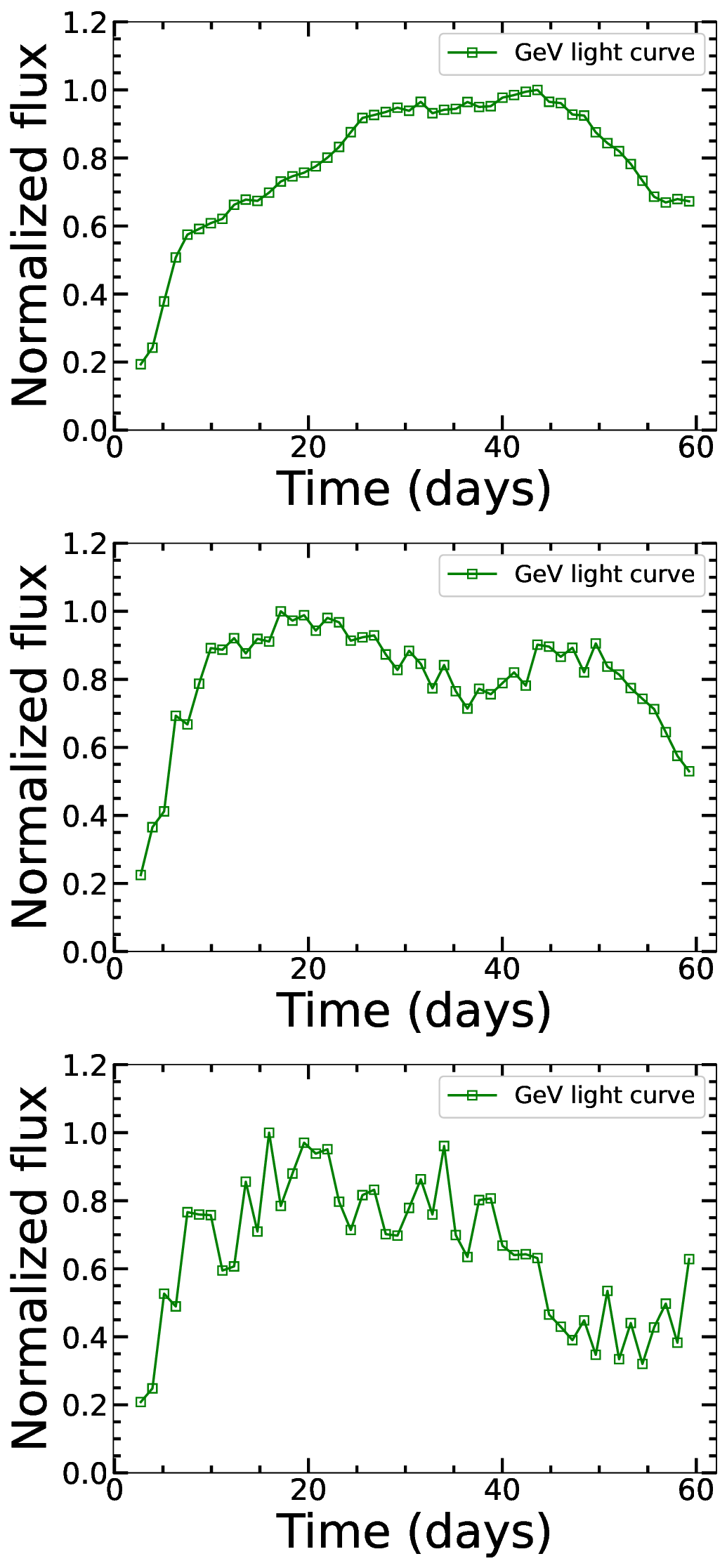}
\caption{GeV light curves with low ($\sim$1-2\%) (\textbf{top panel}), medium ($\sim$10-12\%) (\textbf{middle panel}) and high ($\sim$25-30\%) level of magnetic fluctuations (\textbf{bottom panel}).
\label{fig:EC_and_B_comparison}}
\end{figure}

Optical-GeV correlation studies of large samples of blazars have been carried out by several authors \citep[e.g.,][]{Liodakis.et.al.2019, Majumder.et.al.2019, Rajput.et.al.2020}. Those studies, collectively, may be used to identify statistically significant generalized properties of blazar variability at individual wave bands and their interrelation. For example, most of the studies indicate that the optical and GeV variability are strongly correlated with no time lag. In a small fraction of cases, they have found significant non-zero time lags or orphan flares \citep[e.g.,][]{ori13, cha13a,cha13b,Liodakis.et.al.2019, Rajput.et.al.2020}. Comparison of our model with the above observed results will provide crucial constraints on our model parameters. We simulate light curves at multiple wave bands using a range of parameter values. In order to identify the part of the parameter space that is consistent with the observed results, we cross-correlate the simulated light curve pairs, compute the power spectral density (PSD) of the light curves, and compare with similar results from observed data. The results are given below.

\subsection{Variability Amplitude and Time-scale} \label{subsec:variability}

Light curves displayed in Figure \ref{fig:LC_SED_all_type} left panels show different variability amplitudes and time-scales. The corresponding SEDs for the above light curves are presented in Figure \ref{fig:LC_SED_all_type} right panels. In our model, the short time-scale fluctuations of the synchrotron emission are dominated by the small fluctuations of the magnetic field among the cells in the emission region. 

On the other hand, light curves dominated by the IC process are not directly affected by the same, although those often exhibit short-timescale variability. In the GeV light curves of many blazars, viz. PKS 1510-089 \citep{Saito.et.al.2013}, 3C 454.3 \citep{Britto.et.al.2016}, 3C 273 \citep{Rani.et.al.2013}, S5 0716+714 \citep{Magic.Collaboration.2018} significant short time-scale fluctuations ($\sim$hours) are present. The above blazars are either LSP or ISP type, which means the GeV emission is dominated by the EC process with small or negligible contribution from the SSC component. For the latter, the smallest fluctuations in the light curve may be due to the magnetic field because the seed photons are generated by synchrotron radiation. But the EC process has no such dependence. In our model, the short time-scale fluctuations in the EC light curves are also due to the small-scale fluctuations of the magnetic field through the equipartition of energy between the magnetic field and particles. The equipartition condition in blazar jets has been discussed by other authors \citep[e.g.,][]{Cerruti.et.al.2013, der15, hu17}. We used equipartition to calculate the energy of the particles in the emission region, which can drive small-scale fluctuations of emission. Figure \ref{fig:EC_and_B_comparison} shows the GeV light curves of blazars, in which EC is the dominating inverse-Compton process. We have used $\sim 1\%$, $10\%$ and $25\%$ fluctuations in the magnetic field for variations demonstrated in Figure \ref{fig:EC_and_B_comparison} (top to bottom, respectively). Along with the equal energy distribution, we also tried the partition of energy in different ratios (1:2, 1:5, 5:1, etc.) between the magnetic field and particles, and a similar trend of relation between fluctuations of EC-dominated light curves and the magnetic field has been observed in all cases. Therefore, we conclude that equipartition is a crucial aspect of the energetics of blazar jet emission to be able to reproduce short time-scale small-amplitude fluctuations of the GeV emission of a large fraction of blazars, in which the $\gamma$-rays are generated primarily through the EC process.

\begin{figure}
\includegraphics[width=\columnwidth]{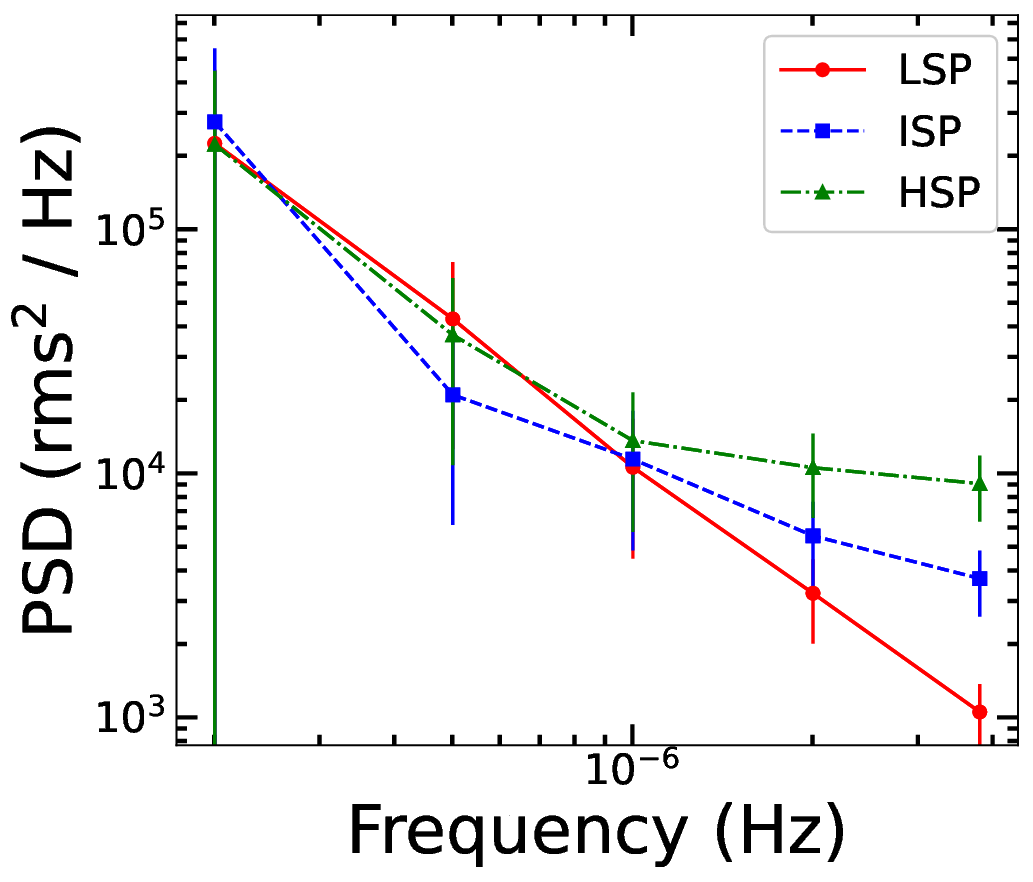}
\caption{PSD of GeV light curves (shown in Figure \ref{fig:LC_SED_all_type}) simulated from our model. For LSP type blazars, a smaller amplitude of short time-scale fluctuation of the magnetic field has been used compared to ISP and HSP type blazars. 
\label{fig:PSD_comparison}}
\end{figure}

In the three sets of light curves shown in Figure \ref{fig:LC_SED_all_type}, we can observe different types of variability in different wavebands. For instance, the optical light curve in Fig \ref{fig:LC_SED_all_type} top left panel shows a smaller variability amplitude at shorter time-scales compared to the same in other panels. This is due to different amounts of small-scale variations included in the amplitude of the magnetic field. The amplitude of fluctuations in the magnetic field used while generating the light curves shown in the top, middle and bottom panels of Figure \ref{fig:LC_SED_all_type} are $\sim$1\%, 10\%, and 25\%, respectively, which is suitable for reproducing the full range of observed synchrotron variability in blazars at $\sim$hr-sub-day timescales \citep[e.g.,][]{Bonning.et.al.2012, Liodakis.et.al.2019}. For the GeV light curves, the difference in variability amplitude at different time-scales is due to the difference in the dominant radiation process. We have computed the typical value of the fractional rms variability amplitude of the simulated light curves with a range of $\gamma_{max}$ values and that from the Fermi-LAT GeV light curves of two representative blazars. Those are shown in Table \ref{tab:std}. Those values indicate that the variability amplitude is significantly larger at longer timescales in the simulated GeV light curve shown in the top left panel in Fig. \ref{fig:LC_SED_all_type}, whereas the same in the middle left and bottom left panels in Fig. \ref{fig:LC_SED_all_type} contain significant variability at both longer and shorter time-scales. $\gamma$-ray light curves of LSP blazars observed by Fermi-LAT often exhibit a similar nature of GeV variability as seen in the top left panel in Fig. \ref{fig:LC_SED_all_type} \citep{Fernandes.et.al.2020} while for HSP blazars fluctuations of the Fermi-LAT light curves are similar to those in the middle left and bottom left panels \citep{Carnerero.et.al.2017}, which is also consistent with values shown in Table \ref{tab:std}.
\begin{table}
\centering
\caption{Fractional rms Variability Amplitude ($F_{var}$)}
\label{tab:std}
\begin{tabular}{lcc}
\hline
Data & $F_{var}$ & $F_{var}$ \\
 & 5-10 days time-scale & 50-100 days time-scale \\
\hline
\hline
LSP (simulated) & 0.05 & 0.18 \\
ISP (simulated) & 0.12 & 0.19 \\
HSP (simulated) & 0.16 & 0.17 \\
\hline
\hline
FSRQ (observed) & 0.11 & 0.26 \\
BL Lac (observed) & 0.26 & 0.36 \\
\hline
\end{tabular}
\end{table}

We compute the power spectral density (PSD) of the light curves, which shows the amplitude of variability at a range of time-scales in the Fourier domain. The PSDs indicate that within the small range of frequencies spanned by those, the light curves are of red noise nature, i.e., smaller amplitude of fluctuations at shorter time-scales, which is a ubiquitous feature of multi-wavelength blazar variability \citep[e.g.,][]{Ritaban.Chatterjee.2008,Ritaban.Chatterjee.2012,Abdo.2010.b, pininti.et.al.2023}. Other authors have also obtained variability of red noise nature in light curves simulated from different kinds of time-dependent leptonic blazar emission models \citep{Thiersen.et.al.2022, Thiersen.et.al.2024, Finke.Becker.2014, Ding.et.al.2024, Chen.et.al.2016}. Furthermore, in Figure \ref{fig:PSD_comparison}, it can be observed that the PSD of the light curve containing a smaller amplitude of the fluctuating component of the magnetic field (Figure \ref{fig:LC_SED_all_type} top left panel) matches well, at lower frequencies, with that of the light curve containing a higher amplitude of the same. However, at higher frequencies, the power is approximately one order of magnitude smaller for the former. This is expected as the short time-scale fluctuations for the light curves shown in the top left panel of Figure \ref{fig:LC_SED_all_type} are smaller compared to the other two. We note that the above discussion is limited within the short range of frequencies spanned by the simulated light curves. If a broadband PSD needs to be computed, then a second PSD at the higher frequencies from a light curve with a shorter time step may be appended to the above PSD. It has been shown by several authors e.g., \citet{Bhattacharyya.et.al.2020} that blazar PSDs are stationary, i.e., the shape tends to remain unchanged with time. Therefore, PSDs covering different ranges of time-scales obtained from different epochs of observations may be appended to each other to construct a broadband PSD.

\subsection{Inter-band Correlation Study} \label{subsec:correlation}

GeV and optical variability in large samples of blazars have exhibited a strong correlation with zero time lag in most cases \citep[e.g.,][]{Liodakis.et.al.2019}. In a few cases, soft lag (optical lagging GeV) or hard lag (GeV lagging optical) has been found. In the cases with strong GeV-optical correlation with zero time delay, the optical flux vs GeV flux plots on a log scale show the same correlation with a slope of $\sim$1.0 for approximately half of the FSRQ-type blazars, and it is $\sim$2.0 for the rest, while that in BL Lac-type objects tends to favor a slope $\sim$2.0. This trend is similar for LSP, ISP and HSP-type blazars \citep{Liodakis.et.al.2019}, i.e., the slope tends to be $\sim$1.0 for almost half and $\sim$2.0 for the rest of the LSPs, while it $\sim$2.0 for most ISPs and HSPs.

\begin{figure}
\includegraphics[width=\columnwidth]{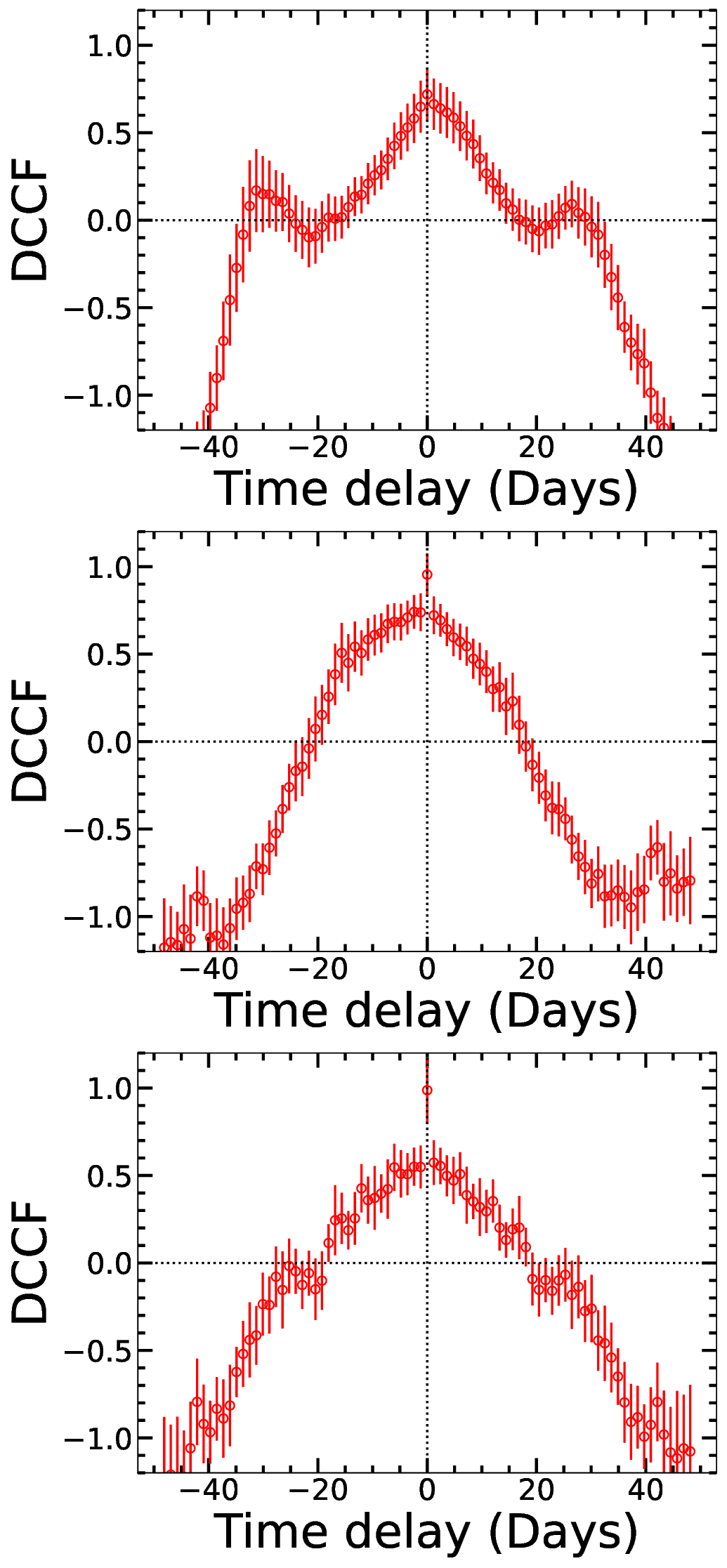}
\caption{Discrete cross-correlation functions of the optical and GeV light curves shown in Figure \ref{fig:LC_SED_all_type}. \textbf{Top} panel shows the DCCF for the LSP type blazars, \textbf{middle} and \textbf{bottom} panel show the same for the ISP and HSP blazars, respectively. In all the cases, we find a strong correlation with time lag consistent with zero.
\label{fig:DCF_all}}
\end{figure}

\begin{figure*}
\includegraphics[width=\textwidth]{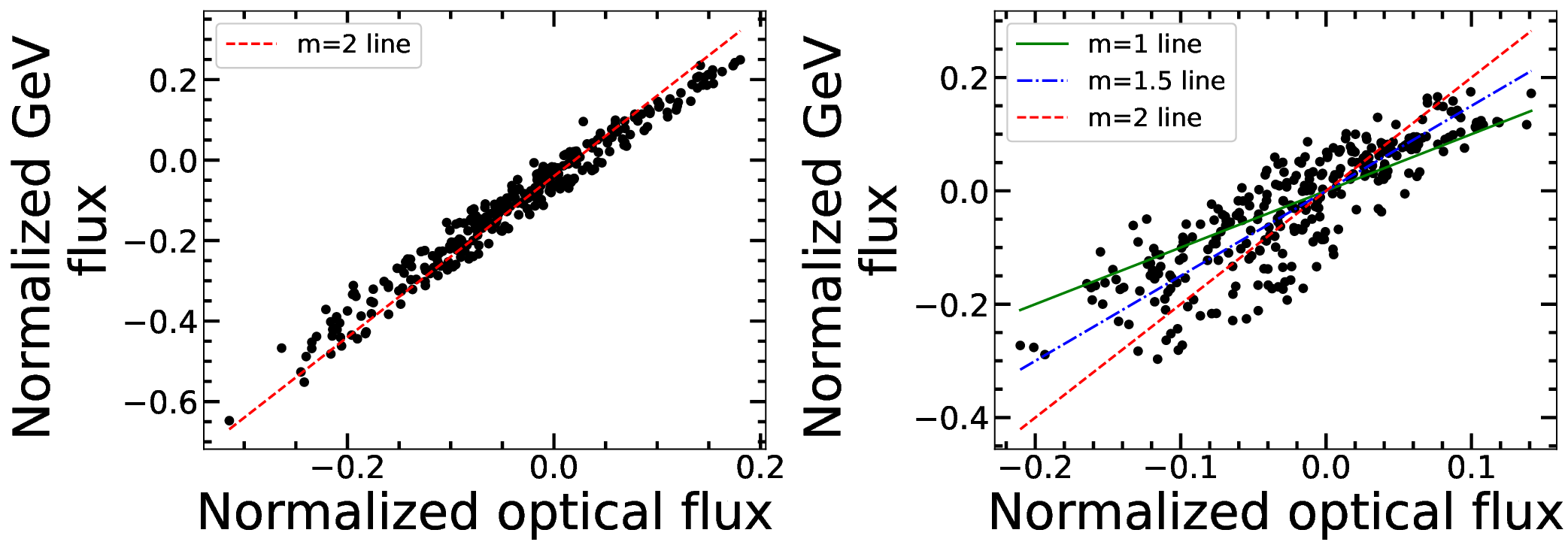}
\caption{Optical-GeV flux comparison for BL Lac type objects (\textbf{left panel}) and FSRQs (\textbf{right panel}), and best-fit lines of different values of slope. The light curves are simulated with a range of values for the relevant parameters as shown in Table \ref{tab:parameter_space}.
\label{fig:m_values}}
\end{figure*}

\begin{figure*}
\includegraphics[width=\textwidth]{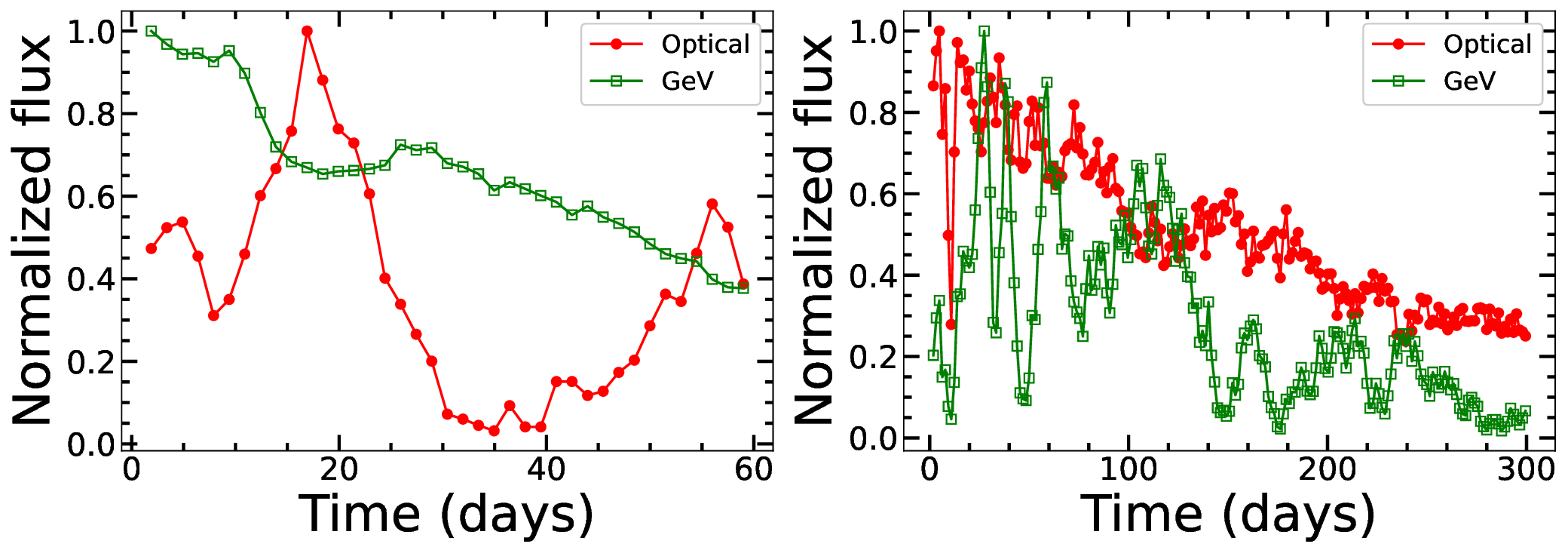}
\caption{Light curves containing orphan flares at optical (\textbf{left panel}) and GeV (\textbf{right panel}) wave band.
\label{fig:Orphan_flares}}
\end{figure*}

\subsubsection{Cross-Correlation} \label{subsubsec:CCF}

To compare our model with the above observed results, we cross-correlated our simulated optical and GeV light curves, and the results have been presented in Figure \ref{fig:DCF_all}. We find that for all LSP, ISP and HSP blazars, the optical-GeV variability is strongly correlated with a time lag consistent with zero in all cases.

Furthermore, we have plotted the normalized optical-GeV flux for BL Lac and FSRQ-type objects in Figure \ref{fig:m_values}. Both synchrotron and EC emission are proportional to the number of electrons, while the SSC emission is proportional to the square of the latter \citep[e.g.,][]{L.Maraschi.1992, G.Ghisellini.1998, Ritaban.Chatterjee.2012}. Therefore, we expect that for BL Lac type objects the slope of the optical vs GeV flux on a log scale will be $\sim$2.0. On the other hand, for FSRQs we expect the GeV emission to be dominated by EC, although there will be some contribution from the SSC process as well. Therefore, the slope may be between 1.0 and 2.0. In the right panel in Figure \ref{fig:m_values}, we can see that the line with a slope of $\sim$1.5 matches best with the data points.

\subsubsection{Orphan Flares} \label{subsubsec:orphan}

In a small fraction of cases, orphan flares have been observed in the simultaneous GeV-optical variability of blazars \citep[e.g.,][]{ori13, cha13a,cha13b,Liodakis.et.al.2019, Rajput.et.al.2020}, i.e., an outburst in one of the wave bands with no significant counterpart in the other. While our simulated multi-wavelength light curves exhibit strong GeV-optical correlation in most cases, it is illuminating to search for conditions that can generate such orphan flares in our model. We find that an orphan flare in the optical band may be generated if there are fluctuations in the turbulent magnetic field (B$_{\rm tur}$) but no or little corresponding variation present in the total magnetic field (B$_{\rm tot}$). Figure \ref{fig:Orphan_flares} left panel shows the simulated optical and GeV light curves where we can see that the optical light curve has an orphan flare, which has no counterpart in the GeV band. In our model, if the angle between the magnetic field and the observer's line of sight is large ($\gtrsim 70^\circ$) then large angular fluctuations will result in prominent fluctuations of B$_\perp$ but smaller variation in B$_\parallel$. Since synchrotron radiation depends on B$_\perp$ but EC does not, the above-mentioned scenario will give rise to optical orphan flares for FSRQs, where the GeV emission is due to the EC process.

\begin{figure*}
\includegraphics[width=\textwidth]{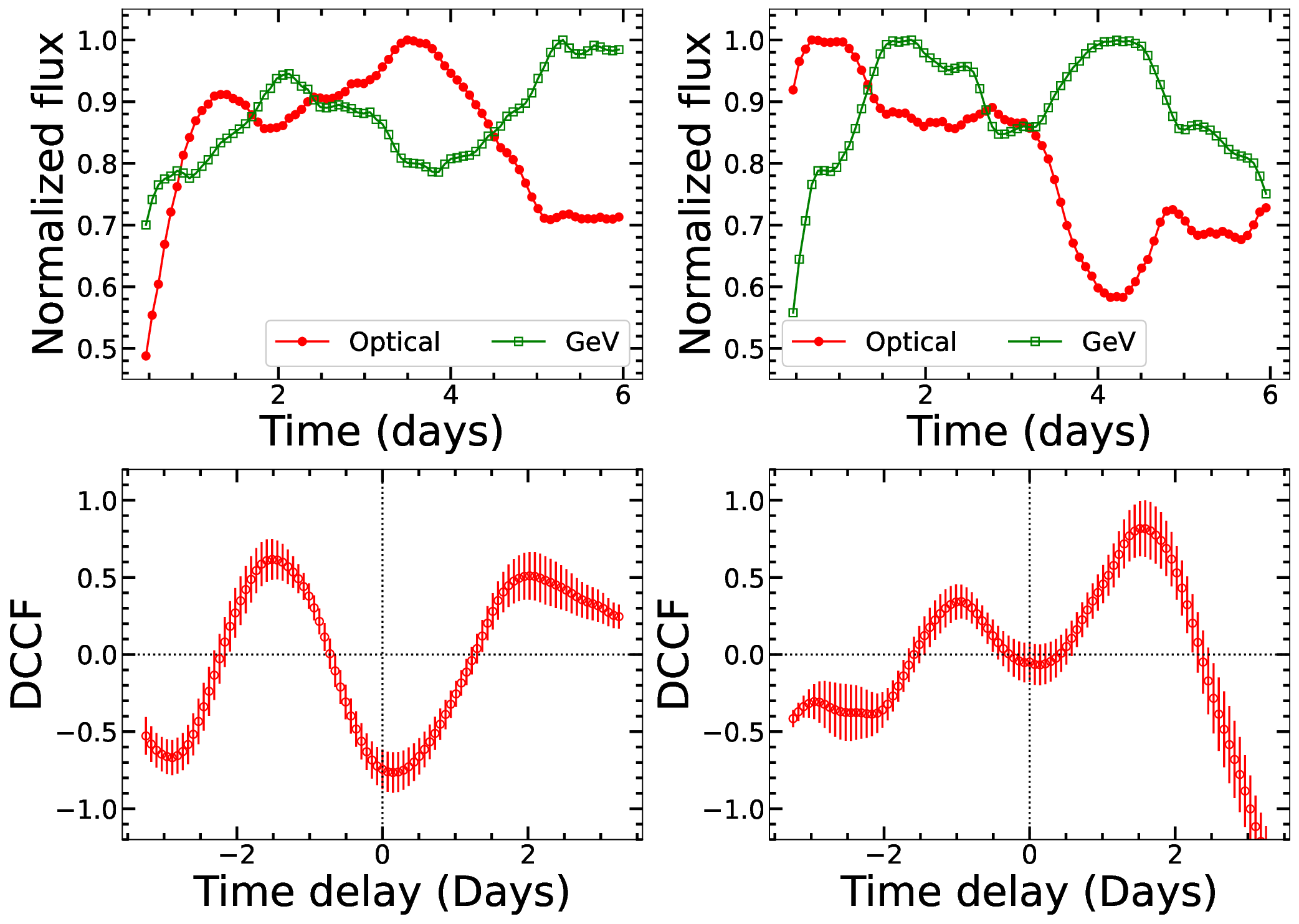}
\caption{Optical-GeV light curves with non-zero lags. \textbf{Left} panel shows the light curves and corresponding discrete cross-correlation function indicating optical variations lagging those at GeV energies. \textbf{Right} panel shows the same with GeV variations lagging the optical.
\label{fig:lags_and_DCFs}}
\end{figure*}

On the other hand, we implement a dependence between $\gamma_{max}$ and magnetic field, e.g., $\gamma_{max} \propto \left(\frac{B_\parallel}{B_{total}}\right)^2$. Such parameterization is motivated by the idea that a magnetic field aligned parallel to the shock normal results in a more efficient energization of emitting particles in the jet by a moving shock front \citep[e.g.,][]{Alan.P.Marscher.2014, Pierson.Romani.2019}. As a result, large fluctuations in the parallel magnetic field will result in variations of $\gamma_{max}$. Such fluctuations can cause an outburst in the GeV band. If the blazar is LSP-type, then there will be similar fluctuations in the optical light curves, too. However, if the blazar is ISP-type, then large flares corresponding to the GeV outburst may be observed in the UV or soft X-ray band while being absent or small at the optical wavelengths. In the latter class of blazars, the synchrotron peak is around the UV frequencies ($\sim$$10^{15}$ Hz) and the fluctuation in $\gamma_{max}$ affects the emission near the peak much more strongly than that at the optical. This may be observed as an orphan flare at the GeV band without any counterpart in the optical band, as shown in Figure \ref{fig:Orphan_flares} right panel.  

\subsubsection{Light Curves with Lags} \label{subsubsec:lags}

\begin{figure}
\includegraphics[width=\columnwidth]{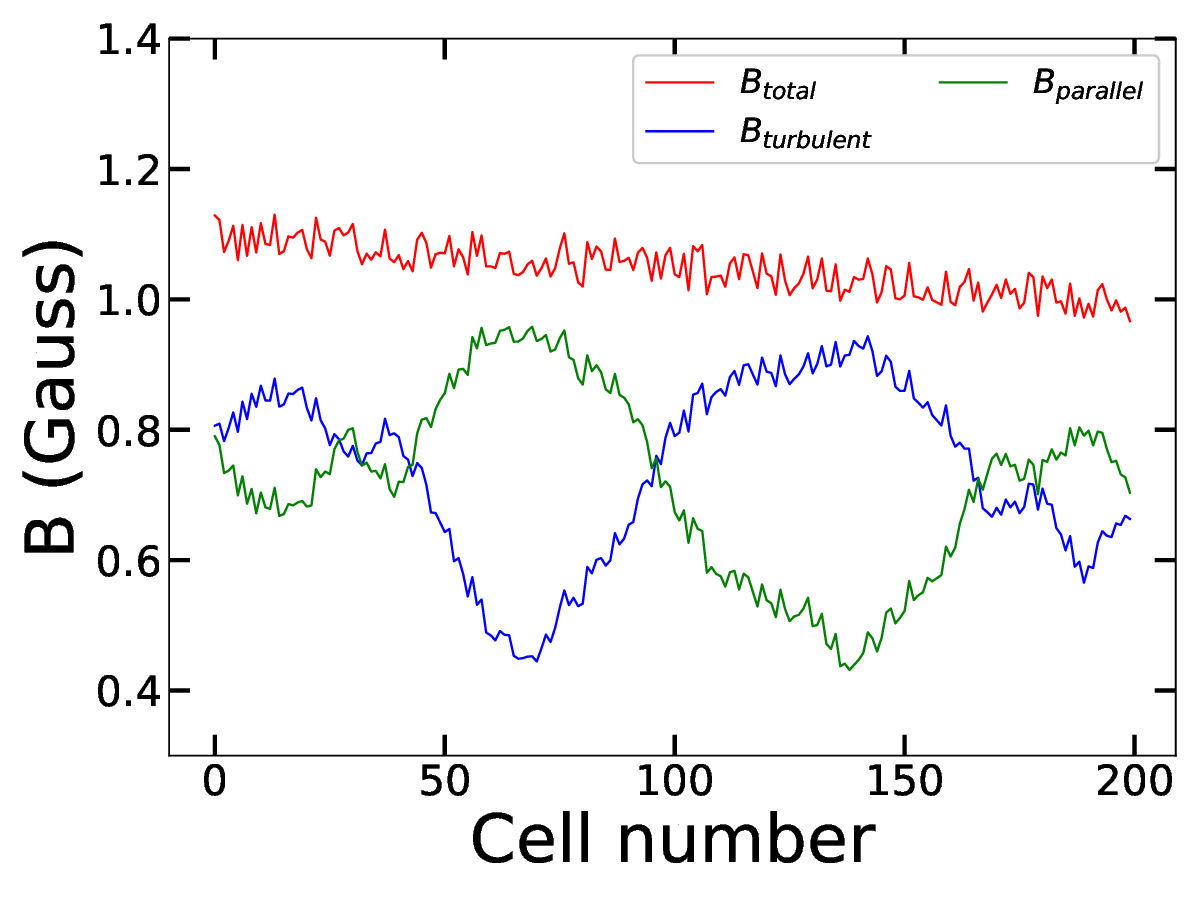}
\caption{A sample configuration of the magnetic field that may be able to generate a significant lag between the variations in the optical and GeV bands. We can observe that the components $B_\perp$ and $B_\parallel$ have an approximately sinusoidal variation with a phase difference.
\label{fig:Mock_B}}
\end{figure}

We have simulated many light curves and in a few cases we have found lags between the optical and GeV variability. Two of those results are shown in Figure \ref{fig:lags_and_DCFs} top panels. The corresponding discrete cross-correlation functions (DCCF) are shown in the bottom panels of Figure \ref{fig:lags_and_DCFs}. We can see that in the left panel, the peak ($\sim$0.9) of the DCCF is at a negative value of time delay, which, in our convention, indicates a soft lag. On the other hand, the DCCF shown in the right panel indicates a hard lag with peak correlation coefficient $\sim$0.8. We have observed that if the angular variation of the magnetic field is low ($\sim 0.1^{\circ} $ per cell) then there is no lag and if the variation is high ($\sim 1^{\circ} $ per cell) then lags (both hard and soft) are present between the optical and GeV variations, which may be as high as a few tens of days. In our model, if there is a comparable level of fluctuations in B$_\parallel$ and B$_\perp$ (Figure \ref{fig:Mock_B}) then both the optical and GeV emissions vary. Since the EC GeV emission depends on $\gamma_{max}$ while synchrotron optical emission depends on $B_\perp$, due to the sine and cosine dependence of the components of the magnetic field (given by Equations 3 and 4) the variations may have a significant lag. Either of the light curves may lead, but as there are other dependencies as well on the emission at those two bands, e.g., on the small spatial fluctuations of the magnetic field, one of the two peaks of the cross-correlation dominates over the other, and that determines whether the optical variations lag those in the GeV band or \textit{vice versa}.

\section{Summary and Discussion} \label{sec:discussion}

In this paper, we have developed a computational model in which the variable multi-wavelength emission from a blazar jet is calculated. The emission region consists of multiple cells with individual magnetic fields and electron distributions. The electrons are instantly energized due to a passing shock and subsequently cool through nonthermal radiation, e.g., synchrotron and IC scattering. The electron energy distribution in individual cells evolves with time, independent of the neighboring cells.

For IC scattering, seed photons from the BLR and torus (external Compton, EC), and synchrotron emission from the jet itself (synchrotron self-Compton, SSC) are considered. For the calculation of the SSC emission, our model does take into account the light-travel-time effects along the jet axis. However, doing so along the transverse direction, i.e., the direction perpendicular to the jet axis, is beyond the scope of this paper, and will be included in a future work. As discussed by, e.g., \citet{Graff.et.al.2008, Joshi.Bottcher.2011}, for a highly collimated jet emission region, having a radius much smaller than the length, considering light travel only along the jet axis is a reasonable assumption. We are primarily considering the observed synchrotron optical vs EC GeV emission variability for comparison with the model results in this work, in which the inaccuracies due to not accounting for light travel along the lateral direction should be minimal.

We examined the accuracy of the above calculation by comparing the resultant SED with observations of blazars. We also found that the nature of the SED changes as expected due to changes in the magnetic field, peak of the electron energy distribution, and the amount of seed photons available for the external Compton process. We also found that, within the short range of frequencies spanned by the simulated light curves, the variability at multiple wave bands follows a red noise power spectrum, i.e., the amplitude of variability is larger at longer timescales as found in observed light curves of blazars. Whether the PSD may be satisfactorily represented by a simple power-law or a change of slope above a certain Fourier frequency is required depends on the relative amplitude of the shorter-timescale variability, which is driven by the spatial fluctuation of the magnetic field in our model. We note that a reliable fit of the PSD in order to obtain any robust inference about the above is possible only if the simulated light curves spanned a larger range of time-scales.

Weeks-to-years timescale variability of blazar emission is supposedly due to the energization of emitting electrons due to the motion of a shock wave down the jet, while radiative cooling of emitting particles is responsible for the sub-day to days timescale variability \citep{Marscher.Gear.1985, Alan.P.Marscher.2014, Bottcher.Baring.2019}. Other models have also been successful in explaining multi-wavelength variabilities at different timescales, such as relativistic magnetic reconnection powering the jet emission \citep{Lorenzo.Sironi.2015}, emitting plasma in the jet following a helical path or jet precession to generate months-to-years timescale variability \citep{Ostorero.L.et.al.2004}, stochastic fluctuations of magnetic field and particle density \citep{ORiordan.et.al.2017}, efficiency of acceleration at shocks or reconnection sites varying with time to generate fluctuations as days to weeks timescale \citep{Kirk.J.et.al.1998, Bottcher.Chiang.2002}, etc. However, the origin of the shortest ($\sim$hr) time-scale fluctuations is not established. In our model, such variability is due to the spatial fluctuation of the magnetic field. The magnetic field directly affects synchrotron radiation that makes up the lower-energy peak of blazar SEDs spanning radio to optical frequencies, sometimes extending to UV or X-rays. We found that to reproduce the short-timescale variability of the observed synchrotron emission in blazars, the required fluctuations of the magnetic field are in the range $1-2\%$ to as large as $25-30\%$ in some cases.  

However, the higher-energy emission, e.g., in the GeV band, is due to IC scattering. For example, in LSP blazars, GeV emission is due to the IC up-scattering of external photons, which is not affected directly by the fluctuation of the magnetic field. Therefore, the origin of the short-timescale variability in those cases was unclear. A critical result we found is that such variability may be due to the change in electron energy driven by the equipartition of energy between the magnetic field and particles.

We focus on the comparison of the simulated GeV and optical variability from our model with that in observations because detailed correlation studies of long-term and well-sampled light curves at those two bands of a large number of blazars are available in the literature. The GeV and optical variability obtained from the scenario adopted in our model are strongly correlated with no significant time lag in most cases, as determined from the above observational studies. In a small fraction of cases, orphan flares in the optical or GeV band or optical-GeV correlation with a significant time delay are observed. We found that those may be reproduced in certain special conditions related to the orientation of the magnetic field and its variation among the cells. Such effects are due to the fact that synchrotron emission is affected by the B$_\perp$ component while B$_{\rm tot}$ is used in the equipartition calculation. The fine-tuning required in order to reproduce the orphan flares or non-zero time lags indicates that those occurrences are relatively uncommon, which is consistent with observations.   

\section*{Acknowledgements}

The authors wish to thank the referee, Prof. Markus B\"ottcher, whose comments and suggestions have greatly improved the draft.
We thank Aritra Kundu for help with the usage of the older version of the code. We thank IUCAA for their hospitality and usage of their facilities during our stay at different times as part of the university associateship program. AB acknowledges financial support from the UGC-NET fellowship. RC thanks ISRO for support under the \textit{AstroSat} archival data utilization program, Presidency University for support under the Faculty Research and Professional Development (FRPDF) Grant, and acknowledges financial support from ANRF through a SURE grant (File No. SUR/2022/001503) and an ARG grant (ANRF/ARG/2025/003315/PS).

\section*{Data Availability}

No observational data have been analyzed in this work. The observational results used for comparison with the theoretical model are from published papers, which have been cited as appropriate.



\bibliographystyle{mnras}
\bibliography{main} 








\bsp	
\label{lastpage}
\end{document}